\documentclass[aps,twocolumn,showpacs,prl]{revtex4}  

\usepackage{graphicx}
\usepackage{mathrsfs}
\usepackage{amsmath}
\usepackage{dcolumn}

\def\beq#1{\begin{equation}\label{#1}}
\def\eeq{\end{equation}}

\begin{document}

\title{Optimal Stochastic Enhancement of Photoionization}
\author{Kamal P. Singh and Jan M. Rost}
\affiliation{Max Planck Institute for the Physics of Complex Systems, 
  N\"othnitzer Stra{\ss}e 38, D-01187 Dresden, Germany}
\date{\today}

\begin{abstract}\noindent

 The effect of noise on the  nonlinear photoionization of an atom due to
 a femtosecond pulse is investigated in the framework of the stochastic 
 Schr\"odinger equation. A modest amount of white noise results in an
 enhancement of the net ionization yield by several orders of
 magnitude, giving rise to a form of quantum stochastic resonance. We
 demonstrate that this effect is preserved if the white noise is replaced by 
 broadband chaotic light.

\end{abstract}


\maketitle
\parindent = 0.4cm

The interplay between noise and the
nonlinear response of a physical system described by classical
mechanics has led to intriguing effects.  A paradigm is stochastic
resonance (SR), whereby a nonlinear system responds more strongly to
coherent subthreshold driving in the presence of an appropriate amount
of noise \cite{Moss,Gamat,Buchl}.  Despite prominent demonstrations of
classical SR in ample variety, only a few quantum mechanical
examples have been studied possibly due to subtle features of the
quantum evolution \cite{Buchl}.  The concept of SR in the
\emph{quantum} domain was originally suggested in the context of
two-level conductance switching of driven mesoscopic metals
\cite{QSR}.  This proposition sparked a number of theoretical studies
on quantum stochastic resonance in double-quantum well structures
\cite{Makri}, the so-called incoherent and coherent SR in driven
dissipative spin-boson systems \cite{Grifoni}, in the bistable
photon-field dynamics in micromaser \cite{mMaser}, and in the electron
shelving in a single ion \cite{Huelga}.  Experiments using NMR
spectroscopy of water \cite{H2O}, and a very recent one in the
radiofrequency controlled nanomechanical beam oscillator \cite{Badzey},
have established the properties of quantum SR in two-level systems (TLS).
These studies are mostly restricted to the quantum analog of the classical 
double-well dynamics.  Yet, they have provided valuable insight into the
noise-induced dynamics of quantum systems.

Following quite a different path of research, there is a rapidly
growing activity in the general area of controlling quantum phenomena
\cite{Rabitz}.  A common approach to exercise the control exploits the
\emph{non-perturbative} interplay between a purposefully designed optical
field from a laser and the target quantum system, such as an atom or
molecule \cite{Zewail}. A fundamental control goal is the
manipulation of atomic and molecular interaction to steer the quantum
system towards a desired state \cite{Rabitz,Zewail}. An accurate
knowledge of the effect of noise on quantum systems
would be very helpful to achieve full control. Their response
to noise has been rarely studied so far \cite{kenny}. One may even
wonder whether the presence of noise offers new possibilities of quantum 
control. 

Here, we will demonstrate the existence of a stochastic
resonance-like effect in a generic quantum situation beyond
the two-level systems. 
For this purpose, we consider a quantum system having a finite binding
potential with mixed, discrete and continuous spectrum which
is coupled to two external forces.
First, a nonresonant coherent optical driving and, second an incoherent
perturbation which may result from some form of environment.
These situations are sufficiently general to be achieved in a
variety of quantum systems such as in nuclear
motion in diatomic molecules, in Josephson junction devices, 
and in active single-electron atoms.

Let us concentrate on the latter example in the form of the simplest 
single electron atom, i.e., the hydrogen atom. Due to the application 
of a linearly polarized laser field $F(t)$,
the electron dynamics is effectively confined to one dimension along
the polarization axis. The Hamiltonian for such a simplified
(yet reliable \cite{Eberly}) description of the hydrogen atom, which 
is here also perturbed by a stochastic force $\xi(t)$,
reads as (atomic units, $\hbar=m=e=1$, are used unless stated otherwise)
 \begin{equation}  
   H(x,t) = \frac{\hat{p}^2}{2} + V(x) + x \lbrace F(t) + \xi (t) \rbrace,
   \label{eqn:eqn1}
 \end{equation}
where $x$ is the position of the electron and $\hat{p}=-i\;\partial/\partial x $ 
is the momentum operator. The potential is approximated 
by a non-singular Coulomb-like form \mbox{$V(x)=-1/\sqrt{x^2 + a^2}$}. 
Such a soft-core potential with parameter $a$ has been routinely employed
to study atomic dynamics in strong laser fields \cite{mpi}. 
It successfully describes many experimental features of multiphoton 
or tunnel ionization \cite{Eberly}, and the observation of the plateau in
the higher harmonic generation spectra \cite{mpi}. The external 
perturbations (last term in Eq.~(1)) is \mbox{dipole-coupled} to the atom. 
The laser field is a nonresonant femtosecond pulse (duration 20 optical periods)
described as, \mbox{$ F(t) = f(t) F_0\sin(\omega t+\delta)$}.
Here \emph{f(t)} defines a smooth pulse envelope with $F_0$ and
$\omega$ denoting peak amplitude and angular frequency, respectively.
The noise term $\xi(t)$ is a zero-mean $\langle \xi(t) \rangle = 0$,
Gaussian white noise with autocorrelation function
 \begin{equation}   
   \langle \xi(t)\xi(t')\rangle = 2D\;\delta (t-t'),
   \label{eqn:eqn2}
 \end{equation}                  
and intensity D \cite{book}.

Due to the stochastic nature of the Hamiltonian, the quantum evolution
is nondeterministic. Thus an averaging over a large number of realizations
of the stochastic force is required in order to produce a statistically
meaningful solution of the time-dependent stochastic Schr\"odinger
equation
 \begin{equation}    
   i\; \frac{\partial \Psi(x,t)}{\partial t} = H(x,t)\; \Psi(x,t).
\label{eqn:eqn3}
\end{equation}
For a given realization, the numerical solution of the  Schr\"odinger equation
amounts to propagating the initial wave function $\vert \Psi_0 \rangle$ using the
 infinitesimal short-time stochastic propagator,
 \mbox{$U_\xi (\Delta t) = \exp\left( -i\int_{t}^{t+\Delta t} H(x,t) dt\right)$}. One can compute $U_\xi (\Delta t)$ using the standard FFT split-operator
 algorithm \cite{SOper}, with the stochastic integration in the exponential 
 interpreted in the Stratonovitch sense \cite{book}. Successive 
 applications of the propagator $U_\xi (\Delta t)$ advance $\vert \Psi_0
 \rangle$ forward in time. Note that the initial state $\vert \Psi_0 \rangle$
 is the \emph{ground state} of the system having a binding energy of 
 $I_b = -0.5$~a.u.. This is obtained by the imaginary-time relaxation
 method for $a^2=2$ \cite{Eberly}.
 To avoid parasitic reflections of the wavefunction from the grid boundary,
 we employ an absorbing boundary \cite{SOper}. The observable, such as the
 ionization flux leaking in the continuum on one side, is defined as 
 $ J_R(x_R,t) = Re\lbrack \Psi^\ast \: \hat{p} \: \Psi \rbrack_{x_R} $,
 where $x_R$ is a distant point (typically 500 a.u.) near the absorbing
 boundary. The ionization rate is integrated
 over a sufficiently long time interval to obtain the corresponding
 total ionization probability, \mbox{$ P = \int_{0}^{\infty} J_R(x_R,t) dt $}.

First, we will discuss the response of the atom interacting
with a short but strong laser pulse only. It produces (nonlinear) ionization
of the atom which is most easily understood, especially in the time domain,
with the picture of tunneling ionization. Ionization flux is produced close
to those times when the potential is maximally bent down  by the 
dipole-coupled laser field.
This is illustrated in Fig.~1 with
the temporal evolution of the
ionization flux for a 20 cycle laser pulse (shown in the top parts of
Fig.~1) with two different peak amplitudes $F_0=0.05$~a.u. [Fig.~1(a)] 
and $F_0=0.02$~a.u. [Fig.~1(b)].
 \begin{figure}[t]   
    \includegraphics[width=.85\columnwidth]{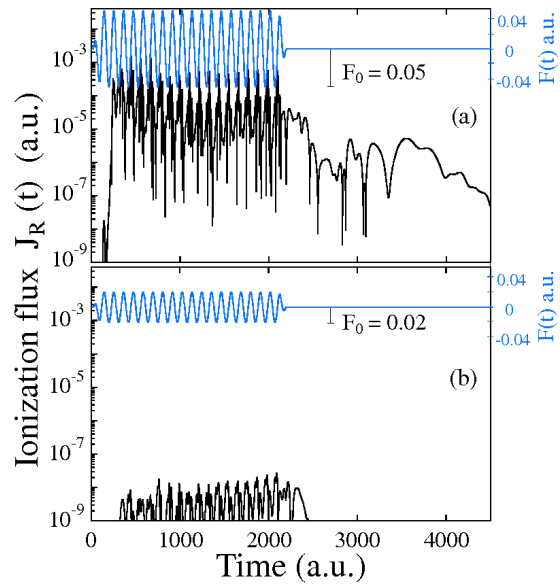}
 \caption{Nonlinear ionization flux $J_R(t)$ (lower part of plots) induced 
 by a 20 optical cycles laser pulse shown in the respective upper plots with
 $\omega=0.057$, $\delta=0$ and peak amplitude, $F_0=0.05$ (a), 
 and $F_0=0.02$ (b).
 The pulse envelope $f(t)$ is unity except single-cycle sinusoidal rising and
 falling edges. The threshold for over barrier 
 ionization is $F_{th}=0.067$~a.u..}
\end{figure}
 Time-resolved ionization peaks separated by the optical period ($2\pi
 / \omega$) are clearly visible for both peak field amplitudes. 
 In addition, $J_{R}(t)$ shows a complex interference pattern due to
 the modulated Coulomb barrier in Fig.~1(a). However, quite strikingly, 
 if  $F_0$ is reduced to 0.02 a.u., 
 the ionization collapses by around five orders of magnitude as shown
 in Fig.~1(b). One can conclude that the photoionization dynamics is highly
 nonlinear, and in particular it exhibits a form of ``threshold''
 dynamics where the threshold is created by the condition for over
 barrier ionization.  
 
 Almost nothing is known about the quantum
 analog of SR in such ``threshold-crossing'' quantum devices
 \cite{Gamat}.  Naturally, in the context of SR the question arises
 if noise can recover the strong periodic ionization flux for the
 ``weak'' laser pulse. To answer this question, we show the ionization flux 
 averaged over 50 realizations in Fig.~2(a) when a small amount of
 noise is added to the weak laser pulse ($F_0=0.02$).
 Note that the noise is switched on for the same time interval as the
 20 cycle laser pulse. One can see that for very small noise
 amplitude $\sqrt{D}=0.00024$, the periodic structure in
 atomic ionization gets amplified by one order of magnitude as
 compared to the case of the laser pulse alone [compare Fig.~1(b)
 with Fig.~2(a)]. However, one might ask if such an enhancement
 could be due to the noise alone.
 Contrary to the coherent excitation, the noise alone produces
 an, on average, featureless ionization profile [see Fig.~2].
 Here the noise causes ionization from the atomic ground state, which is
 different from the studied inization of the Rydberg atoms \cite{NoiseIonRyd}.
 This purely stochastic ionization for the feeble noise
 $\sqrt{D}=0.00024$ is considerably smaller,
 than the corresponding case of laser pulse with noise [Fig.~2(a)].
 Hence, the observed net enhancement can be attributed 
 to a nonlinear quantum interaction between coherent pulse
 and noise, which is also beyond the response to a simple quantum addition
 of the individual external fields.

As the noise level is further increased, we observe an enhancement
of the periodic ionization profile by many orders of magnitude
as shown in Fig.~2(b)-(c). However, the increase in 
noise level also causes the stochastic ionization curve to rise
rapidly. Eventually, for strong noise case, the coherent structures
tend to wash out and the noise dominates the ionization [Fig.~2(c)].
This suggests the existence of an optimum ratio between the noise
and laser amplitudes which leads to a maximum ionization enhancement.
\begin{figure}[t]
  \begin{center}
    \includegraphics[width=.80\columnwidth]{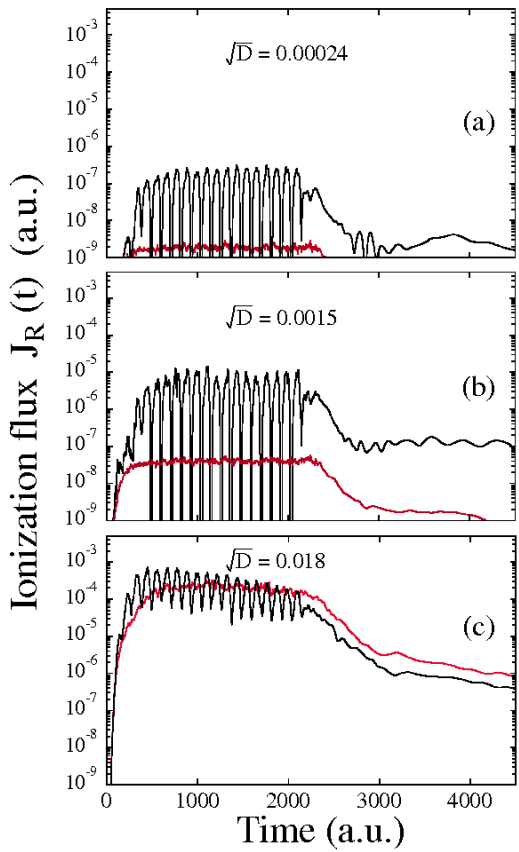}
    \caption{
 Ionisation flux for a weak laser pulse $F_0=0.02$,
 with three values of noise amplitude $\sqrt{D}$ (a) $0.00024$, 
 (b) $0.0015$, and (c) $0.018$. Background featureless curves
 show the corresponding purely noise-driven ($F_0=0$) flux.
 The flux is averaged over 50 realizations.}
\label{fig:fig2}
   \end{center}
\end{figure}

\parindent = 0.4cm
To quantify the quantum stochastic enhancement we define
the enhancement factor 
\begin{equation}            
\eta = \frac{P_{s+n} - P_0}{P_0} 
\label{eqn:eqn4}
\end{equation}
 with $P_{0} = P_s + P_n $. Here $P_{s+n}$ denotes the average ionization
 probability (IP) due to the presence of the laser pulse with noise, 
 $P_s$ and $P_n$ are the individual IP for laser pulse and noise,
 respectively. Although this is different compared to the
 quantifiers commonly used \cite{Gamat,Buchl}, $\eta$ is more suitable for our case.
 One can verify that a zero value of $\eta$ corresponds to the case 
 when either the laser pulse ($P_s \gg P_n$) or the noise ($P_s \ll P_n$) dominates.
 Furthermore, $\eta$ characterizes a truly nonlinear quantum interplay 
 as it also vanishes if we assume a ``linear'' response as a sum of individual IP, 
 $P_{s+n} = P_s + P_n $. In Fig.~3 we have plotted the enhancement factor
 $\eta$ versus the noise amplitude.
 It exhibits a sharp rise, followed by a maximum at a certain value of the
 noise (point B), and then a gradual fall off. 
 It is worth mentioning that only a modest noise to laser ratio 
 ($\sqrt{D_{opt}}/F_0 = 0.075$)
 is required to reach the optimum enhancement 
 (here $\eta_{max} = 36$), as indicated by a typical optimal pulse shape
 in the inset of Fig.~3. 
\begin{figure}[b]    
\includegraphics[width=.95\columnwidth]{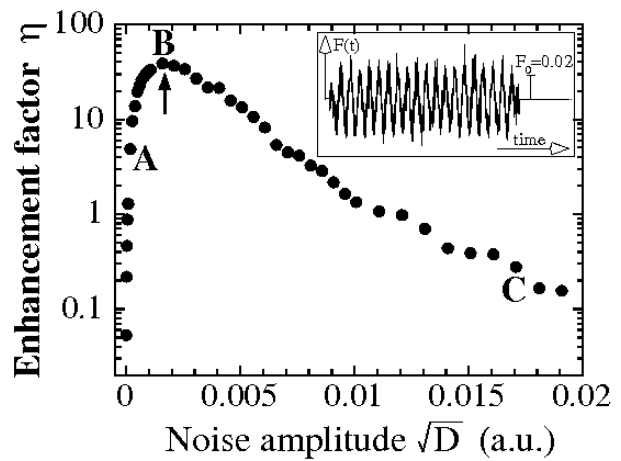}
\caption{The enhancement in photoionization due to quantum SR.
 The points marked A-C correspond to the  noise amplitudes of
 Fig.~2(a)-(c), respectively. Inset: A typical optimal pulse
 for point B with $\sqrt{D_{opt}}=0.0015$, $F_0=0.02$.}
\label{fig:fig3}
\end{figure}

 The enhancement in photoionization can be understood by a simple 
 two-step mechanism. 
 First, starting from the ground state, the atom absorbs energy from the
 noise leading to a Boltzman-like population of energy levels. 
 In a second step, the laser field causes ionization from the electron 
 wavefunction Boltzmann distributed over many (excited) states.  This
 simple picture is indeed verified by separating the laser 
 irradiation and the noise input in time.  
 A sequential application of the noise
 followed by the laser pulse leaves the SR curve almost invariant.
 On the contrary, by reversing the sequence, i.e., laser pulse first and then
 noise, destroys the SR. Note that due to the level structure of the 
 atom (many bound states and continuum)
 there is no relevant noise induced time-scale as known from 
 two-level systems. Thus, the optimum enhancement does not show the
 characteristic synchronization between coherent and noise induced
 time-scales as in the TLS \cite{Gamat,Buchl}.  
 Indeed, in some classical systems SR has been shown to exist
 without any explicit synchronization requirement \cite{Bezrukov}.
 Atomic ionization under a driving laser
 field provides the quantum analog to synchronization free SR. 

 It is worth mentioning that the  features presented here are robust
 with respect to the choice of parameters. For example, we have 
 observed the quantum SR for pulses lasting from a few cycles to few
 hundred cycles, and for $\omega$ ranging  over more than one order of
 magnitude from infrared to near UV frequencies.
\begin{figure}[t]         
  \begin{center}
    \includegraphics[width=.95\columnwidth]{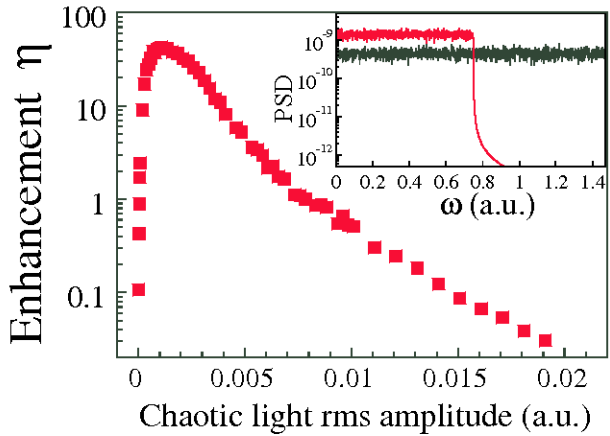}
     \caption{
Enhancement induced by a broadband chaotic light. 
The peak amplitude and frequency of the 20 cycle laser pulse are $F_0=0.02$ and $\omega=0.057$,
respectively. The bandwidth of chaotic light is $\Delta\omega=0.75$ with
central frequency $\omega_0 =0.375$. 
Inset: power spectral density (PSD) of the chaotic light 
compared to the one for the white noise.}
\label{fig:fig4}
  \end{center}
\end{figure}

The intensity required for the white noise renders an experimental
test of our prediction quite challenging. However, one could think
of replacing the white noise by a quasi-white noise, i.e., chaotic
light with a finite but broad bandwidth $\Delta\omega$. This can 
be realized by adding a large number (here 1024) of independent
phase-randomized frequency modes. The inset of Fig.~4 shows an 
example of a characteristic spectral density of such a broadband
source having $\Delta\omega=20~eV$ ($0.75~a.u.$).
Modern pulse shaping techniques can generate chaotic light pulses
with a bandwidth as large as 30 eV \cite{chaoLgt,Martn}.
In Fig.~4 we have plotted $\eta$ defined in Eq.~(4) vs the rms 
amplitude of the chaotic light (analogous to the noise amplitude)
for the previously used coherent pulse.
One can see that such chaotic light can preserve the features of
quantum SR, with an almost identical optimum compared to the one for
the white noise case of Fig.~3. This observation may creat new possibilities
in quasi-coherent control schemes of similar quantum systems.

In conclusion, we have demonstrated a new form of quantum 
stochastic resonance in the dynamics of the simplest atomic system
for the first time. This generalized quantum SR leads to a dramatic
enhancement (by several orders of magnitude) in the nonlinear ionization
when a \emph{modest} amount of optimum white noise is added to the
weak few cycle laser pulse. The same effect is also achieved if 
one uses (realizable) broadband chaotic light instead of white noise.
We emphasize that the effet is robust with respect to a range of
experimentally accessible parameters. In addition to substantially
broadening the existing paradigm for quantum SR to generic atomic and 
molecular systems, our results might provide valuable insight into 
the possible role of noise in designing \emph{optimal} quasi-coherent
quantum control schemes \cite{Rabitz,Martn}. Finally, analogous effects are
also expected in other systems such as in photo-fragmentation of anharmonic
diatomic molecules \cite{Chelk,Dunbar}, and in the recently observed
multiphoton transitions in current-biased Josephson devices \cite{squid}, provided a coupling with an incoherent perturbation exists.

 We thank A. Kenfack, W. Peijie, N. Singh, A. Buchleitner, and
 P. H\"anggi for fruitful discussions.

\end{document}